# 1/f NOISE UNDER DRIFT AND THERMAL AGITATION IN SEMICONDUCTOR MATERIALS

FERDINAND GRÜNEIS

*Institute for Applied Stochastic*
*Rudolf von Scholtz Str. 4, 94036 Passau, Germany*
*Email: Ferdinand.Grueneis@t-online.de*

Voss and Clarke observed 1/f noise in the square of Johnson noise across samples in thermal equilibrium without applying a current. We refer to this phenomenon as "thermal 1/f noise". Voss and Clarke suggested spatially correlated temperature fluctuations as an origin of thermal 1/f noise; they also showed that thermal 1/f noise closely matches the 1/f spectrum obtained by passing a current through the sample. An intermittent generation-recombination (g-r) process has recently been introduced to interpret 1/f noise in semiconductors. The square of this intermittent g-r process generates a 1/f noise component which correlates with Voss and Clarke's empirical findings. Traps which intermittently rather than continuously generate g-r pulses are suggested as the origin of 1/f noise under drift and thermal agitation. We see no need to introduce correlated temperature fluctuations or oxide traps with a large distribution of time constants to explain 1/f noise.

## 1. Introduction

Voss and Clarke [1-2] reported measurements in semiconductors and metals which strongly suggest that 1/f noise is an equilibrium phenomenon. They observed 1/f noise in the square of Johnson noise across samples in thermal equilibrium without using any current; this supports referring to this phenomenon as "thermal 1/f noise". Voss and Clarke showed that thermal 1/f noise closely matches the 1/f spectrum obtained by passing a current through the sample. This equivalence is expressed by

$$\frac{S_{I_d}(f)}{I_0^2} = \frac{S_P(f)}{\overline{P}^2} = \frac{C_{1/f}}{f}. \tag{1.1}$$

Herein, $I_0$ is an applied current and $S_{I_d}(f)$ the spectrum of current fluctuations due to drift; $\overline{P}$ is the mean thermal power and $S_P(f)$ the spectrum of thermal power fluctuations being equivalent to fluctuations in the square of Johnson noise. $C_{1/f}$ is a constant related to the number of conduction electrons [3-4]. Beck and Spruit [5] later confirmed Voss and Clarke's results; in addition, they provided a criterion for the detection limit of thermal 1/f noise. Hashiguchi [6] also found that the variance of Johnson noise in metal micro-bridges exhibits a 1/f noise component closely related to resistance fluctuations in current-carrying micro-bridges.

Voss and Clarke suggested spatially correlated temperature fluctuations as the origin of thermal 1/f noise. Under certain conditions, these fluctuations may yield an extended 1/f region. The question remains why 1/f noise under drift is equal to thermal 1/f noise. The theoretical importance of these empirical findings justifies further investigations.

Kleinpennig and de Kuijper [7] investigated a related problem: They showed that the variance of a 1/f noise signal grows logarithmically with the sample duration. Voss and Clarke chose an alternative approach by applying spectral analysis to the variance of Johnson noise. For the following we refer to Voss and Clarke's approach.

In this paper, we present an interpretation of 1/f noise under drift and thermal agitation in semiconductor materials matching Voss and Clarke's empirical findings. Our approach to g-r noise and 1/f noise is based on an intermittent generation-recombination (g-r) process which has recently been introduced as an interpretation of Hooge's relation [8-9]. Extending this approach to thermal noise, we provide a theoretical framework for thermal power fluctuations. The inclusion of 1/f noise into this framework is straightforward providing an interpretation of 1/f noise under drift and thermal agitation. Our approach is an addendum to theoretical and experimental work on noise in semiconductors which is excellently summarized in the overview article by Mitin, Reggiani, and Varani [10] and in the book by Bonani and Ghione [11].

## 2. Fluctuations in semiconductors

In a doped n-type semiconductor, we only consider transitions between the level of traps and the conduction band to produce fluctuations (Fig. 1). For steady-state conditions, $g_0$ and $r_0$ are the generation and recombination probabilities per time unit, respectively, and $N_0$ denotes the mean number of conduction electrons. The number of fully-ionized, shallow donors is $N_D$ and the number of traps in the semiconductor material, $N_T$. For steady-state conditions, the mean number of conduction electrons is $N_0 = N_{T+} + N_D$. For $N_0 \gg 1$, the fluctuations of conduction electrons about mean value $N_0$ attain a normal distribution. Under this condition, the master-equation approach provides [12-13]

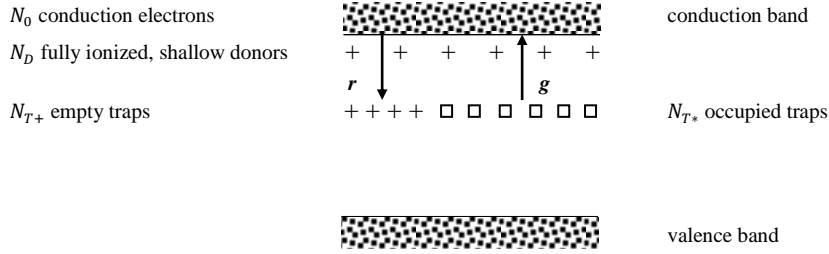

$N_0$ conduction electrons — conduction band
$N_D$ fully ionized, shallow donors
$N_{T+}$ empty traps — $N_{T*}$ occupied traps
valence band

Fig. 1. Electron transitions for an extrinsic n-type semiconductor with traps. Under steady-state conditions, $N_{T+}$ is the number of empty traps and $N_{T*}$ the number of occupied traps. $N_T = N_{T+} + N_{T*}$ is the number of traps which are neutral when occupied. $N_0$ is the number of conduction electrons and $N_D$ the number of fully ionized, shallow donors.

$$g_0 = \overline{\Delta N_{gr}^2}/\tau_{gr} \qquad (2.1)$$

where $\overline{\Delta N_{gr}^2}$ is the mean square fluctuations of conduction electrons contributing to the g-r process; $\tau_{gr}$ is the g-r lifetime. Generations occurring at random obey Poisson statistics implying [14]

$$\overline{\Delta N_{gr}^2} = \overline{N_{gr}} \qquad (2.2)$$

where $\overline{N_{gr}}$ is the mean number of conduction electrons contributing to the g-r process. We define by

$$\eta_{gr} = \frac{\overline{N_{gr}}}{N_0} \qquad (2.3)$$

the fraction of the conduction electrons contributing to the g-r process. In a bulk semiconductor $0 < \eta_{gr} < \frac{1}{2}$. Substituting Eq. (2.2) into Eq. (2.1) leads to

$$g_0 = \overline{N_{gr}} / \tau_{gr}. \qquad (2.4)$$

Here, the generation rate is defined by the parameters of the conduction electrons. Alternatively, by using the parameters of traps, the generation rate is likewise defined by [14]

$$g_0 = N_T/\tau_T. \qquad (2.5)$$

$\tau_T$ represents the time between successive generations at an individual trap. As is shown in [14], fluctuations due to the g-r process can be attributed to a random succession of elementary g-r pulses. This approach is also applied to an intermittent g-r process investigated in the following.

### 2.1. 1/f noise due to the intermittent g-r process

For interpreting 1/f noise in semiconductors an intermittent g-r process has been introduced [8-9]. This section summarizes the spectral features of the intermittent g-r process as far as they are of relevance for the subsequent sections. Just as in blinking single-quantum dots and other materials [15-17], the generation process of an individual trap in a semiconductor is thought to be intermitted by a gating process with an off-time $\tau_{off}$ and an on-time $\tau_{on}$ (Fig. 2.a). An intermittent g-r process is thus obtained (Fig. 2.b).

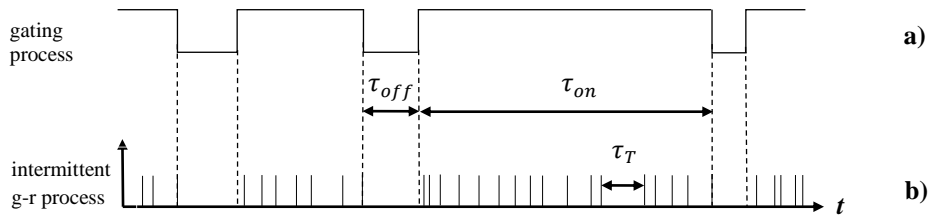

Fig. 2. a) The two-state process with states $\tau_{off}$ (= intermission) and $\tau_{on}$ (= lifetime of a cluster) which gates the g-r process of an individual trap. b) Successive generations due to the intermittent g-r process. The time between successive generations is $\tau_T$. Each spike triggers a g-r current pulse.

The off-times are assumed to be exponentially distributed. As in quantum dots and other materials, the on-times follow a power-law distribution such as $1/t^{\mu_{on}}$. This leads to a finite and random number of spikes in so-called clusters. These number fluctuations can be described by a cluster-size distribution $q_n$ that follows the power-law distribution of on-times by $q_n \propto 1/n^{\mu_{on}}$. Denoting the mean number of spikes in a cluster by $\overline{N_c} = \sum n\, q_n$ the mean on-time is obtained by (Fig. 2.b)

$$\tau_{on} = \overline{N_c}\, \tau_T. \qquad (2.6)$$

The intermittent g-r process in Fig. 2.b represents generations resulting from an individual trap; each spike triggers an elementary g-r current pulse $h_d(t)$ as is seen in Fig. 3.a. The power spectrum due to all $N_T$ traps exhibits reduced g-r noise and a 1/f noise component [8-9]

$$S_{I_d}^{1/f}(f) = g_0 |\overline{H_d(0)}|^2 \frac{\alpha_{im}}{(f\tau_T)^b}. \qquad (2.7)$$

The parameters of on-off intermittency (indicated by subscript *im*) are comprised in the coefficient

$$\alpha_{im} = 2C(r) \frac{\tau_T}{\tau_{on}+\tau_{off}}. \qquad (2.8)$$

$C(r) \approx 0.3[r/(3+r)]^2$ is a pre-factor which depends on the normalized off-time $r = \tau_{off}/\tau_T$.

After generation, an electron remains in the conduction band for the g-r lifetime time $\tau_{gr}$ before it recombines into an arbitrary empty trap. As is seen in Fig. 3.a, this leads to a g-r current pulse $h_d(t)$; the subscript $d$ indicates that electrons drift along an applied electric field.

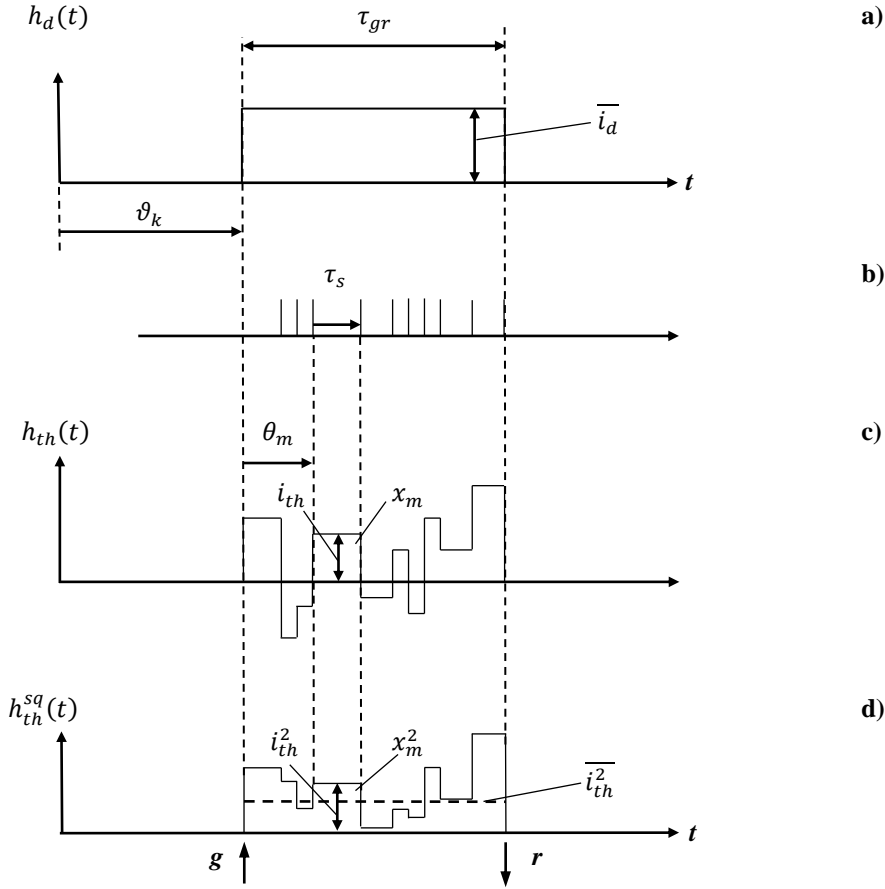

Fig. 3. a) A g-r current pulse $h_d(t)$ with g-r lifetime time $\tau_{gr}$ and amplitude $\overline{i_d} = ev_d/L$; the time point of occurrence is $\vartheta_k$. $e$ is the elementary charge and $L$ the length of the resistor. Under an applied electric field $E_0$, the mean drift velocity of an electron is $v_d = \mu E_0$ with $\mu$ being the electron mobility. b) The spikes indicate the random time points of electron-phonon scattering during the g-r process; $\tau_s$ is the time between collisions. c) The current fluctuations of an electron during the g-r lifetime time $\tau_{gr}$ caused by thermal agitation; this leads to a g-r pulse $h_{th}(t)$. d) The corresponding thermal current fluctuations $h_{th}^{sq}(t)$ for squared amplitudes $i_{th}^2$; after averaging, this leads to a g-r pulse with amplitude $\overline{i_{th}^2} = (e/L)^2 \overline{v_{th}^2}$ and lifetime $\tau_{gr}$ (dotted line).

The Fourier transform of the g-r current pulse $h_d(t)$ is denoted $H_d(f)$. Considering that the g-r lifetime $\tau_{gr}$ is exponentially distributed we find

$$\overline{|H_d(f)|^2} = \frac{(\tau_{gr}\,\overline{i_d})^2}{1+(2\pi f \tau_{gr})^2}. \tag{2.9}$$

Substituting this into Eq. (2.7) and considering that an applied current is equal to a mean drift current $I_0 = \overline{I_d} = N_0\,\overline{i_d}$, we obtain

$$S_{I_d}^{1/f}(f) = I_0^2 \frac{\eta_{gr}^2}{g_0} \frac{\alpha_{im}}{(f\tau_T)^b}. \tag{2.10}$$

Applying Eq. (2.5), 1/f noise can be given an alternative form of Hooge's relation

$$S_{I_d}^{1/f}(f) = I_0^2 \frac{\eta_{gr}^2 \tau_T}{N_T} \frac{\alpha_{im}}{(f\tau_T)^b} \tag{2.11}$$

relating 1/f noise to the number of traps rather than to the number conduction electrons as was defined by Hooge [3-4]. The normalized 1/f noise component is given by

$$\frac{S_{I_d}^{1/f}(f)}{I_0^2} = C_{1/f} \frac{\tau_T}{(f\tau_T)^b} \tag{2.12}$$

where

$$C_{1/f} = \frac{\alpha_{im}\,\eta_{gr}^2}{N_T} \tag{2.13}$$

is a coefficient comprising the parameters of the intermittent g-r process.

## 3. Thermal current and power fluctuations

We model thermal fluctuations on a simple electron gas [18]. $N_0$ is the mean number of free electrons contributing to the conduction mechanism. Electron-phonon scattering is characterized by a single time constant $\tau_s$. Johnson's measurements of thermal noise [19] led Nyquist [20] to the power spectrum of thermal current fluctuations $I_{th}(t)$ by

$$S_{I_{th}}(f) = \frac{4kT}{R} \frac{1}{[1+(2\pi f \tau_s)^2]}. \tag{3.1}$$

$T$ is the temperature of the resistor $R$. Nyquist's relation applies up to a cut-off frequency corresponding to the electron-phonon scattering time $\tau_s$. Quantum effects will be neglected throughout the paper. The short-circuit current fluctuations of the resistor $I_{th}(t)$ are Gaussian distributed with mean value $\overline{I_{th}} = 0$. The mean square of thermal current fluctuations is determined by

$$\overline{I_{th}^2} = \int_0^\infty S_{I_{th}}(f)df = \frac{kT}{R\tau_s}. \tag{3.2}$$

Introducing the effective bandwidth of thermal noise by [21]

$$\Delta f_{th} \equiv \frac{\int_0^\infty S_{I_{th}}(f)df}{S_{I_{th}}(0)} = 1/4\tau_s \tag{3.3}$$

Eq. (3.2) can also be expressed by

$$\overline{I_{th}^2} = \frac{4kT}{R}\Delta f_{th} = S_{I_{th}}(0)\Delta f_{th}. \tag{3.4}$$

This is the mean square of the thermal current fluctuations provided by the resistor $R$. At room temperature [18], the electron-phonon scattering time $\tau_s \approx 10^{-12} - 10^{-14}$ s leading to an effective bandwidth of about $10^{12} - 10^{14}$ Hz.

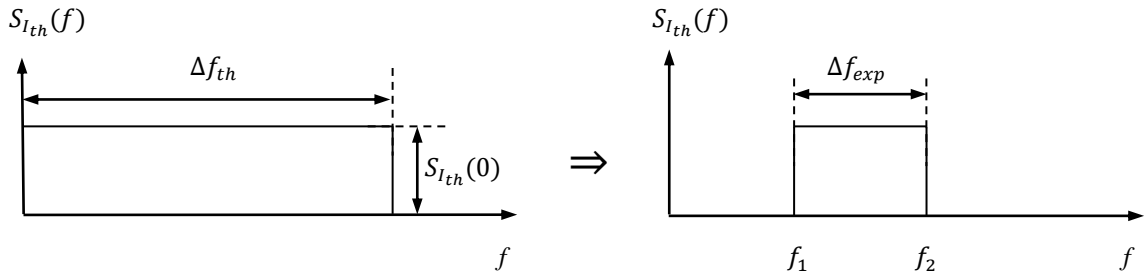

Fig. 4. Left figure: the effective bandwidth $\Delta f_{th}$ of total thermal noise provided by the resistor $R$. Right figure: for an experiment measuring the noise of the resistor $R$, the bandwidth is reduced to $\Delta f_{exp}$.

For an experiment measuring the noise of the resistor $R$, however, the bandwidth is determined by the lower and upper frequency limit $f_1$ and $f_2$ of the pre-amplifier. This leads to $\Delta f_{exp} = f_2 - f_1$ which replaces $\Delta f_{th}$ (Fig. 4). A bandwidth of $\Delta f_{exp} \approx 10^5 - 10^6$ Hz has been reported [1, 5, 6]. In any case $\Delta f_{exp} \ll \Delta f_{th}$, i.e. the band-pass filter of the pre-amplifier is much smaller than the effective bandwidth of thermal noise. This justifies neglecting quantum effects which come into play for frequencies above $10^{13}$ Hz at room temperature.

### 3.1. *Thermal power noise*

The thermal power is also a fluctuating quantity; it derives from the square law detection of thermal noise by

$$P(t) = R\, I_{th}^2(t). \tag{3.5}$$

Herein, $P(t)$ is thermal power noise. The power spectra of the thermal power and of the squared thermal fluctuations are related by

$$S_P(f) = R^2\, S_{I_{th}^2}(f). \tag{3.6}$$

This justifies using the term "thermal power noise" as a synonym for "fluctuations of squared thermal noise". The mean value of thermal power is

$$\overline{P} = R\,\overline{I_{th}^2} = 4kT\Delta f_{th}. \tag{3.7}$$

In combination with Eq. (3.3) we get

$$\overline{P} = \frac{kT}{\tau_s} = \frac{\overline{E}}{\tau_s}. \tag{3.8}$$

The harmonic oscillator has two degrees of freedom each contributing with $\frac{1}{2}kT$ to $\overline{E} = kT$. This is the mean thermal energy which is transferred from the heat bath to the semiconductor during the electron-phonon scattering time $\tau_s$. The mean square fluctuations of the harmonic oscillator are found in standard textbooks on thermodynamic by [22]

$$\overline{\Delta E^2} = \overline{E}^2. \tag{3.9}$$

According to Eq. (3.8), this also applies to thermal power fluctuations resulting in

$$\overline{\Delta P^2} = \overline{P}^2 = (4kT\Delta f_{th})^2. \tag{3.10}$$

This yields a white power spectrum comprising the effective bandwidth $\Delta f_{th}$ by

$$S_P(f) = \frac{\overline{\Delta P^2}}{\Delta f_{th}} = (4kT)^2 \Delta f_{th}. \tag{3.11}$$

Applying Eq. (3.6) leads to

$$S_{I_{th}^2}(f) = (4kT/R)^2 \Delta f_{th} \tag{3.12}$$

coinciding with a result derived by Beck and Spruit [5] in a different way.

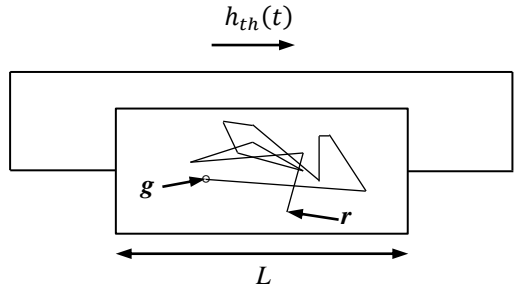

Fig. 5. Illustration of the random walk of an electron in thermal equilibrium due to electron-phonon scattering during g-r lifetime $\tau_{gr}$ (**g** = generation and **r** = recombination of an electron). The time between collisions is $\tau_s$. $L$ is the length of the semiconductor. During g-r lifetime an electron induces a short circuit current pulse train $h_{th}(t)$ seen in Fig. 3.c.

### 4. The spectral contributions of the g-r process to thermal power noise

For the following, we consider the random walk of an electron during g-r lifetime in thermal equilibrium (Fig. 5). Under this condition, $h_d(t) = 0$ and we are left with the thermal g-r current pulse (Fig. 3.c)

$$h_{th}(t) = \sum_{m=1}^{N_s} x_m(t - \theta_m). \tag{4.1}$$

$x_m(t)$ is a rectangular pulse and $\theta_m$ is the occurrence time of the m-th current pulse. $N_s$ is the number of scattering processes during the g-r lifetime. Such g-r pulses occurring at random cause g-r current fluctuations described by

$$I_{th}^{gr}(t) = \sum_{k=-\infty}^{+\infty} h_{th}(t - \vartheta_k). \tag{4.2}$$

$\vartheta_k$ is the time point of occurrence (Fig. 3.a). The contribution of the g-r process to thermal power noise is

$$P_{gr}(t) = R \left\{I_{th}^{gr}(t)\right\}^2. \tag{4.3}$$

Squaring Eq. (4.2) yields

$$\left\{I_{th}^{gr}(t)\right\}^2 = \sum\sum_{k \neq l} h_{th}(t - \vartheta_k) h_{th}(t - \vartheta_l) + \sum_{k=-\infty}^{+\infty} h_{th}^2(t - \vartheta_k). \tag{4.4}$$

The power spectrum (dimension A$^4$/Hz) of this signal has been derived by Papoulis [23]. Aside a direct current (DC) component, Papoulis provides

$$S_{P_{gr}}(f) = S_{P_{gr}}^{Gauss}(f) + S_{P_{gr}}^{sq}(f). \tag{4.5}$$

The first term is due to the double sum in Eq. (4.4); it is the Gaussian distributed spectral contribution of the g-r process to thermal power noise. The second term is the spectral contribution due to the sum at the far right side of Eq. (4.4); this sum represents g-r pulses $h_{th}^2(t)$ occurring at random with mean rate $g_0$. Denoting the Fourier transform of the g-r pulse $h_{th}^2(t)$ by $H_{th}^{sq}(f)$ and applying Carson's theorem [24], the power spectrum is obtained by

$$S_{P_{gr}}^{sq}(f) = 2 R^2 g_0 \overline{\left|H_{th}^{sq}(f)\right|^2}. \tag{4.6}$$

For $g_0 \gg 1$, as is the case for a bulk semiconductor, Papoulis [23] showed that $S_{P_{gr}}^{Gauss}(f) \gg S_{P_{gr}}^{sq}(f)$; for this reason, an evaluation of Eq. (4.6) is not needed.

### 4.1. Thermal 1/f noise – the spectral contribution of the intermittent g-r process to thermal power noise

For the intermittent g-r process, the g-r pulses $h_{th}^2(t)$ are generated no longer at random but intermittently (Fig. 2.b). As a consequence, Eq. (4.6) splits up into a white noise component (being well below thermal power noise) and into a 1/f noise component (exceeding thermal power noise for frequencies sufficiently low). Replacing the Fourier transform of $h_d(t)$ in Eq. (2.7) by the Fourier transform of $h_{th}^2(t)$, we obtain the 1/f noise component of thermal power noise by

$$S_{P_{gr}}^{1/f}(f) = R^2 g_0 \left|\overline{H_{th}^{sq}(0)}\right|^2 \frac{\alpha_{im}}{(f\tau_T)^b}. \tag{4.7}$$

Substituting Eq. (A.10) and using Eqs. (2.13) and (3.7) this is transformed into

$$S_{P_{gr}}^{1/f}(f) = \overline{P}^2 C_{1/f} \frac{\tau_T}{(f\tau_T)^b}. \tag{4.8}$$

This is thermal 1/f noise which adds to thermal power noise (Eq. (3.11)). Fig. 6 illustrates the corresponding spectral contributions. The corner frequency where a pure 1/f shape ($b = 1$) exceeds the plateau of thermal power noise is

$$f_c = C_{1/f} \Delta f_{th}. \tag{4.9}$$

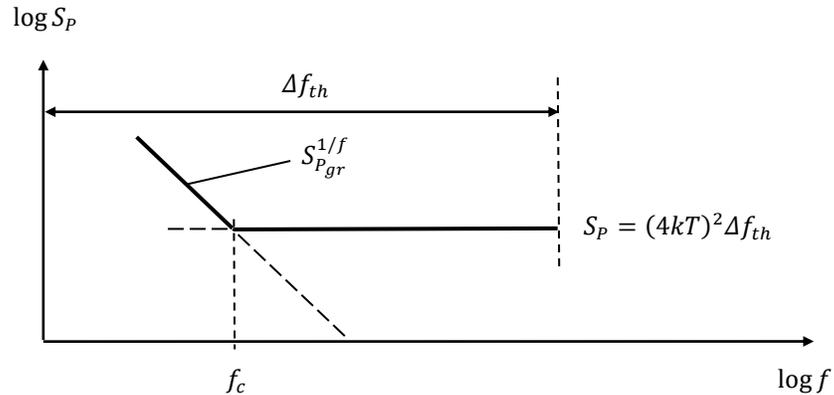

Fig. 6. Thermal 1/f noise in excess to thermal power noise in a log-log scale. $f_c$ is the corner frequency where thermal 1/f noise is equal to thermal power noise.

Comparing Eq. (4.8) with Eq. (2.12), the normalized fluctuations can be expressed by

$$\frac{S_{I_d}^{1/f}(f)}{I_0^2} = \frac{S_{P_{gr}}^{1/f}(f)}{\bar{P}^2} = C_{1/f}\frac{\tau_T}{(f\tau_T)^b} \qquad (4.10)$$

suggesting the equivalence of 1/f noise under drift and under thermal agitation. The common origin of 1/f noise in both cases is the intermittent nature of the g-r process. Eq. (4.10) corresponds to Voss and Clarke's empirical findings comprised in Eq. (1.1) applying also for $b \neq 1$. On the basis of the intermittent g-r process, the constant $C_{1/f}$ is specified in Eq. (2.13).

## 5. Results and Discussions

Voss and Clarke observed 1/f noise in the square of Johnson noise across semiconductor and metal film samples in thermal equilibrium without using any current [1-2]. Their experiments suggest that 1/f noise is a phenomenon that takes place in thermal equilibrium which may justify denoting this phenomenon as "thermal 1/f noise".

In this paper, thermal 1/f noise is interpreted on the basis of an intermittent g-r process. Such an intermittent g-r process has recently been introduced for deriving an alternative form of Hooge's relation [8-9]: by analogy to single quantum dots and other materials [15-17], we assume that the g-r process due to an individual trap is controlled by on-off-states. A power-law distributed on-state leads to an extended 1/f region. We show that the square of the intermittent g-r process generates a 1/f noise component which also obeys an alternative form of Hooge's relation. This concurs with Voss and Clarke's observations that thermal 1/f noise closely matches 1/f noise measured under an applied current.

As an origin of thermal 1/f noise, Voss and Clarke suggest spatially correlated temperature fluctuations, which may give rise to an extended 1/f region. As an alternative interpretation, we suggest an intermittent g-r process exhibiting power-law distributed on-states but uncorrelated temperature fluctuations. Such an intermittent g-r process is shown to be the common origin of 1/f noise under drift and thermal agitation. We favor traps, which do not continuously, but intermittently generate g-r pulses. Such behavior is thought to be due to intermittent phonon scattering, which may be caused by non-linear interactions between phonon modes [25]. Fingerprints for a phonon-induced 1/f noise have also been reported by Mihaila [26].

Based on an idea of Mc Worther [27], oxide traps with a large distribution of time constants are often considered as an interpretation of 1/f noise in semiconductors [28-29]. We do not need such oxide traps with a large distribution of time constants in our model. The origin of 1/f noise is thought to be an intrinsic property of the semiconductor caused by traps which generate g-r pulses intermittently.

1/f noise can be generated in several different ways. For example, Kirton and Uren [30] and Handel [31-32] established theories which explain 1/f noise in semiconductors under drift. It is an open question whether these theories apply also for 1/f noise under thermal agitation matching Voss and Clarke's empirical findings.

## Appendix A. The Fourier transform of the g-r pulse $h_{th}^2(t)$

As is seen in Fig. 3.c, the g-r pulse $h_{th}(t)$ can be expressed by

$$h_{th}(t) = \sum_{m=1}^{N_s} x_m(t - \theta_m) \qquad (A.1)$$

where $\theta_m$ is the occurrence time of the m-th current pulse $x_m(t)$ with lifetime $\tau_s$ and amplitude $i_{th} = ev_{th}/L$; $v_{th}$ is the thermal velocity. The number of scattering processes during g-r lifetime is $N_s$ (Fig. 3.b) yielding

$$\tau_{gr} = \overline{N_s}\,\tau_s. \qquad (A.2)$$

Squaring Eq. (A.1), we obtain

$$h_{th}^2(t) = \sum_{m=1}^{N_s} x_m^2(t - \theta_m) + \sum\sum_{m \neq n} x_m(t - \theta_m)x_n(t - \theta_n). \qquad (A.3)$$

Indicating the Fourier transform by the symbol $\mathcal{F}$, we define $H_{th}^{sq}(f) \equiv \mathcal{F}\{h_{th}^2(t)\}$ and $X_{sq,m}(f) \equiv \mathcal{F}\{x_m^2(t)\}$ leading to

$$H_{th}^{sq}(f) = \sum_{m=1}^{N_s} X_{sq,m}(f) + \sum\sum_{m \neq n} \mathcal{F}\{x_m(t - \theta_m)x_n(t - \theta_n)\}. \qquad (A.4)$$

According to Eq. (4.7), we need $\overline{H_{th}^{sq}(f)}$. Assuming the temperature fluctuations to be independent, the mean over the second term yields

$$\overline{\sum\sum_{m \neq n} \mathcal{F}\{x_m(t - \theta_m)x_n(t - \theta_n)\}} = \sum\sum_{m \neq n} \mathcal{F}\left\{\overline{x_m(t - \theta_m)}\;\overline{x_n(t - \theta_n)}\right\}. \qquad (A.5)$$

Considering that the mean of a current pulse $\overline{x_m(t)} \propto \overline{i_{th}} = 0$, the average over the second term in Eq. (A.4) is zero and we are left with the average over the first term by

$$\overline{H_{th}^{sq}(f)} = \overline{N_s}\,\overline{X_{sq}(f)}. \tag{A.6}$$

$X_{sq}(f)$ is the Fourier transform of a pulse $x_m^2(t)$ which is characterized by the amplitude $i_{th}^2$ and the lifetime $\tau_s$ (Fig. 3.d); observing that $\tau_s$ is exponentially distributed we find

$$\left|\overline{X_{sq}(f)}\right|^2 = \frac{\tau_s^2\,\overline{i_{th}^2}^2}{1+(2\pi f \tau_s)^2} \tag{A.7}$$

and in combination with Eq. (A.2)

$$\left|\overline{H_{th}^{sq}(f)}\right|^2 = \frac{\left(\tau_{gr}\,\overline{i_{th}^2}\right)^2}{1+(2\pi f \tau_s)^2}. \tag{A.8}$$

This shows that $\overline{H_{th}^{sq}(f)}$ can be interpreted as the Fourier transform of a g-r current pulse with a squared amplitude $\overline{i_{th}^2}$ and lifetime $\tau_{gr}$ (dotted line in Fig. 3.d). In context with Eq. (4.7), we evaluate

$$g_0 \left|\overline{H_{th}^{sq}(0)}\right|^2 = \frac{1}{g_0}\left(g_0 \tau_{gr}\,\overline{i_{th}^2}\right)^2 = \frac{1}{g_0}\left(\frac{N_{gr}}{N_0} N_0\,\overline{i_{th}^2}\right)^2. \tag{A.9}$$

Using Eqs. (2.3), (2.5) and (B.5) we end up with

$$g_0 \left|\overline{H_{th}^{sq}(0)}\right|^2 = \frac{\tau_T}{N_T}\left(\eta_{gr}\,\overline{I_{th}^2}\right)^2. \tag{A.10}$$

**Appendix B. The microscopic interpretation of thermal current fluctuations**

The mean square of thermal current fluctuations $\overline{I_{th}^2}$ can also be given a microscopic interpretation: the mean square of a thermal current pulse is (Fig. 3.d)

$$\overline{i_{th}^2} = \left(\frac{e}{L}\right)^2 \overline{v_{th}^2}. \tag{B.1}$$

Applying the equipartition theorem, the velocity fluctuations are obtained by $\frac{1}{2}m_e\overline{v_{th}^2} = \frac{1}{2}kT$ leading to

$$\overline{i_{th}^2} = \left(\frac{e}{L}\right)^2 \frac{kT}{m_e}. \tag{B.2}$$

$m_e$ is the mass of the electron. The resistor $R$ is defined by

$$\frac{1}{R} = \frac{A\,\sigma}{L} = \frac{A}{L}\frac{e^2 n_0 \tau_s}{m_e} \tag{B.3}$$

$A$ being the area of the resistor. Considering also $n_0 = N_0/AL$, we obtain

$$\frac{kT}{R\tau_s} = N_0\,\overline{i_{th}^2}. \tag{B.4}$$

By comparison with Eq. (3.2) we find

$$\overline{I_{th}^2} = N_0\,\overline{i_{th}^2}. \tag{B.5}$$

**Acknowledgement**

The author would like to thank Dr. Barbara Herzberger for proofreading this manuscript.